\documentclass{llncs}

\usepackage[fleqn]{amsmath}
\usepackage[psamsfonts]{amssymb}
\usepackage[final]{graphicx}
\usepackage{braket}

\newfont{\bg}{cmr9 scaled\magstep4}

\title{Voronoi Diagrams and a Numerical Estimation of a Quantum Channel Capacity}
\titlerunning{VD's and a Numerical Estimation of a Quantum Channel Capacity}

\author{
Kimikazu Kato\inst{1,2} \and Mayumi Oto\inst{3} \and Hiroshi Imai\inst{1,4} \and Keiko Imai\inst{5}}
\institute{
Department of Computer Science,University of Tokyo
\and
Nihon Unisys, Ltd.
\and 
Software Engineering Center, Toshiba Corporation
\and
ERATO-SORST Quantum Computation and Information
\and
Department of Information and
System Engineering, Chuo University
}

\def\Tr{{\rm Tr}\;}

\begin{document}
\thispagestyle{plain}
\pagestyle{plain}
\sloppy

\maketitle
\begin{abstract}
 We give a new geometric interpretation of quantum pure states.  Using
 Voronoi diagrams, we reinterpret the structure of the space of pure
 states as a subspace of the quantum state space. In addition to the
 known coincidence of some Voronoi diagrams for one-qubit pure states,
 we will show that even for mixed one-qubit states, as far as sites are
 given as pure states, the Voronoi diagram with respect to some
 distances --- the divergence, the Bures distance, and the Euclidean
 distance --- are all the same.

As to higher level pure quantum states, for the divergence, the
 Fubini-Study distance, and the Bures distance, the coincidence of the
 diagrams still holds, while the coincidence of the diagrams with
 respect to the divergence and the Euclidean distance no longer
 holds. That fact has a significant meaning when we try to apply the
 method used for a numerical estimation of a one-qubit quantum channel
 capacity to a higher level system.
\end{abstract}
\section{Introduction}

The movement of trying to apply quantum mechanics to information
processing has given vast research fields in computer science
\cite{NC00}. Especially among them, the field which pursues the
effectiveness of a quantum communication channel is called quantum
information theory. Some aspect of quantum information theory is to
investigate a kind of distance between two different quantum
states. Depending on the situation, several distances are defined in
quantum states. In quantum information geometry, the structure of those
distances is of great interest \cite{amari00,petz96}.

In classical information geometry, Onishi and Imai
\cite{onishi97,onishi97a} did a computational geometric analysis using a
Voronoi diagram. A Voronoi diagram and a Delaunay triangulation are
defined with respect to the Kullback-Leibler divergence, and are
shown to be the extensions of the Euclidean counterparts. The Voronoi
diagram is computed from an associated potential function instead of a
paraboloid which is used in a Euclidean Voronoi diagram.

In this paper, we extend the Voronoi diagram in classical geometry to
the quantum setting. In quantum information theory, there is a natural
extension of the Kullback-Leibler divergence, which is called a quantum
divergence. We discuss a Voronoi diagram with respect to the quantum
divergence, and analyze its structure.

For pure states in the space of one-qubit quantum states, the authors
showed the coincidence of Voronoi diagrams with respect to some
distances --- the divergence, the Fubini-Study distance, the Bures
distance, the geodesic distance and the Euclidean distance
\cite{kato05}. Here the diagram with respect to the divergence can be
defined by taking a limit of the diagram in mixed states. As an
application of this fact, we introduced a method to compute numerically
the Holevo capacity of a quantum channel \cite{oto04}. The effectiveness
of this method is partially based on the coincidence of the
diagrams. Moreover, also as to the diagrams in mixed states with sites
given as pure states, we found the coincidence of some of them; the
diagrams with respect to the three distances --- the divergence, the
Fubini-Study distance, and the Bures distance --- coincide
\cite{kato06b}.

A natural question that arises after this story is ``What happens in a
higher level system?'' For a higher level system, the diagrams with
respect to the divergence and the Euclidean distance do not coincide
anymore \cite{kato06a}. On the other hand, the diagrams with respect to
the divergence, the Bures distance and the Fubini-Study distance still
coincide for a higher level.

The rest of this paper is organized as follows. In Sect.\
\ref{sec:preliminaries}, we give some definitions and prepare some
mathematical tools. In Sect.\ \ref{sec:one-qubit}, we give a theorem
about a Voronoi diagram for one-qubit quantum states.  In
Sect. \ref{sec:holevo}, we explain a method to compute the Holevo
capacity and its relation to the proven theorem. In Sect.\
\ref{sec:higher-level} and Set.\ \ref{sec:higher-level-bures}, we extend
the discussion of one-qubit Voronoi diagram to a higher level. Lastly we
give a conclusion in Sect.\ \ref{sec:conclusion}. The latest result of
our research is the latter half of Sect.\ \ref{sec:one-qubit} (Theorem
\ref{th:one-qubit-mixed}) and Sect.\ \ref{sec:higher-level-bures}. We
give all the theorems without detailed proofs. The proofs will be given
in the paper being prepared \cite{kato06b}.

\section{Preliminaries}\label{sec:preliminaries}
\subsection{Parameterization of quantum states}
In quantum information theory, a {\em density matrix} is a
representation of some probabilistic distribution of states of
particles. A density matrix is expressed as a complex matrix which
satisfies three conditions: a) It is Hermitian, b) the trace of it is
one, and c) it must be semi-positive definite. We denote by
$\mathcal{S}(\mathbb{C}^d)$ the space of all density matrices of size
$d\times d$. It is called a {\em $d$-level system.}

Especially in a two-level system, which is often called a {\em one-qubit}
system, the conditions above are equivalently expressed as
\begin{equation}
  \label{one-qubit-rho}
 \rho= \left(
 \begin{array}{cc}
 \displaystyle \frac{1+z}{2}& \displaystyle \frac{x-iy}{2} \\
 \displaystyle \frac{x+iy}{2}& \displaystyle \frac{1-z}{2}
 \end{array}
 \right),
 x^2+y^2+z^2 \leq 1,\quad x,y,z\in \mathbb{R}.
\end{equation}
The parameterized matrix correspond to the conditions a) and b), and the
inequality correspond to the condition c). In this case, a density
matrix correspond to a point in a ball. We call it a {\em Bloch ball.}

There have been some attempt to extend this Bloch ball expression to a
higher level system. A matrix which satisfies only first two conditions,
Hermitianness and unity of its trace, is expressed as:
\begin{align} \label{multi-level-rho}
\lefteqn{\rho=}\nonumber\\ 
&{
\left(
\begin{array}{ccccc}
\displaystyle\frac{\xi_1+1}{d} &
\!\!\!\displaystyle\frac{\xi_d-i\xi_{d+1}}{2} &
\cdots & &   
\displaystyle\frac{\xi_{3d-4}-i\xi_{3d-3}}{2} 
\smallskip\\
\displaystyle\frac{\xi_d+i\xi_{d+1}}{2} & 
\!\!\!\displaystyle\frac{\xi_2+1}{d} &
\cdots & & 
\displaystyle\frac{\xi_{5d-8}-i\xi_{5d-7}}{2} 
\smallskip\\
\vdots & & \!\!\!\!\ddots & & \vdots \\
\displaystyle\frac{\xi_{3d-6}+i\xi_{3d-5}}{2} & \cdots& &
\!\!\!\!\!\!\!\displaystyle\frac{\xi_{d-1}+1}{d}& 
\displaystyle\frac{\xi_{d^2-2}-i\xi_{d^2-1}}{2}\smallskip\\
\displaystyle\frac{\xi_{3d-4}+i\xi_{3d-3}}{2} & \cdots& &
\!\!\!\!\!\!\!\displaystyle\frac{\xi_{d^2-2}+i\xi_{d^2-1}}{2} &
\displaystyle
 \frac{-\sum_{i=1}^{d-1}\xi_i+1}{d}
\end{array}
\right)},
\quad \xi_i\in \mathbb{R}.
\end{align}
Actually, any matrix which is Hermitian and whose trace is one is
expressed this way with some adequate $\{\xi_i\}$.  This condition
doesn't contain a consideration for a semi-positivity. To add the
condition for a semi-positivity, it is not simple as in one-qubit case;
actually we have to consider complicated inequalities
\cite{byrd03,kimura03}. Note that this is not the only way to
parameterize all the density matrices, but it is reasonably natural way
because it is natural extension of one-qubit case and has a special
symmetry.

Additionally our interest is a pure state. A pure state is expressed by
a density matrix whose rank is one. A density matrix which is not pure
is called a mixed state. A pure state has a special meaning in quantum
information theory and also has a geometrically special meaning because
it is on the boundary of the convex object. For one-qubit states, the
condition for $\rho$ to be pure is simply expressed as $x^2+y^2+z^2=1$.
This is a surface of the Bloch ball. On the other hand, in general case,
the condition for pureness is again expressed by complicated equations.

\subsection{Some distances and the Holevo capacity}

For two pure states $\rho$ and $\sigma$, the Fubini-Study distance 
$d_{\rm FS}(\rho,\sigma)$ is defined by
\begin{equation}
 \cos d_{\rm FS}(\rho,\sigma)=\sqrt{\Tr(\rho\sigma)},
\qquad 0\leq d_{\rm FS}(\rho,\sigma)\leq\frac{\pi}{2}.
\end{equation}
See Hayashi \cite{hayashi98}.  The Bures distance $d_{\rm B}(\rho,\sigma)$
\cite{bures69} is defined by
\begin{equation}
 d_{\rm B}(\rho,\sigma)=\sqrt{1-\Tr(\rho\sigma)}.
\end{equation}
Moreover, the Bures distance is also defined for mixed states.
When $\rho$ and $\sigma$ are mixed states, their Bures distance is
defined as
\begin{equation}
 d_\mathrm{B}(\rho,\sigma)=\sqrt{1-\Tr \sqrt{\sqrt{\sigma}\rho\sqrt{\sigma}}}.
\end{equation}
Since $\Tr \sqrt{\sqrt{\sigma}\rho\sqrt{\sigma}}=\Tr \rho\sigma$ when
$\rho$ and $\sigma$ are pure, this definition is consistent with the
definition above for pure states.

The quantum divergence is one of measures that show the
difference of two quantum states. The quantum divergence of the two
states $\sigma$ and $\rho$ is defined as
\begin{equation}
  D(\sigma||\rho) = \Tr \sigma (\log \sigma - \log \rho) .
\end{equation}
Note that though this has some distance-like properties, it is not commutative, i.e.~$D(\sigma||\rho)\neq
D(\rho||\sigma)$. The divergence $D(\sigma||\rho)$ is not defined when
$\rho$ does not has a full rank, while $\sigma$ can be non-full
rank. This is because for a non-full rank matrix, a log of zero appears in
the definition of the divergence. However, since $0\log 0$ is naturally
defined as $0$, some eigenvalues of $\sigma$ can be zero.

A quantum channel is the linear transform that maps quantum states to
quantum states. In other words, a linear transform
$\Gamma:M(\mathbb{C};d)\to M(\mathbb{C};d)$ is a quantum channel if
$\Gamma(\mathcal{S}(\mathbb{C}^d))\subset \mathcal{S}(\mathbb{C}^d)$.

The Holevo capacity \cite{holevo98} of this quantum channel is known to
be equal to the maximum divergence from the center to a given point and
the radius of the smallest enclosing ball. The Holevo capacity
$C(\Gamma)$ of a 1-qubit quantum channel $\Gamma$ is defined as
\begin{equation}
   C(\Gamma)= \inf_{\sigma\in \mathcal{S}(\mathbb{C}^d)} 
\sup_{\rho\in \mathcal{S}(\mathbb{C}^d)} D(\Gamma(\sigma)||\Gamma(\rho)).
\end{equation}

\section{Voronoi Diagrams for One-qubit Quantum
 States}
\label{sec:one-qubit}

We define the Voronoi diagrams with respect to the divergences as
follows.

\begin{align}
V_D(v_i)
=\bigcap_{i\neq j}\left\{\sigma | D(\sigma||\rho(v_i))\geq
 D(\sigma||\rho(v_j))\right\},\nonumber\\
 V_{D^*}(v_i)
=\bigcap_{i\neq j}\left\{\sigma | D(\rho(v_i)||\sigma)\geq D(\rho(v_j)||\sigma)\right\}.
\end{align}

 Note the quantum divergence is only defined for the mixed states.
  Actually, while $D(\rho||\sigma)=\Tr\rho(\log\rho - \log\sigma)$ can
  be defined when an eigenvalue of $\rho$ equals $0$ because $0\log 0$
  can be naturally defined as $0$, it is not defined when an eigenvalue
  of $\sigma$ is $0$. Here we show that this Voronoi diagram of mixed
  states can be extended to pure states. In other words, we prove that
  even though the divergence $D(\rho||\sigma)$ can not be defined when
  $\sigma$ is a pure state, the Voronoi edges are naturally extended to
  pure states. In other words, we can define a Voronoi diagram for pure
  states by taking a natural limit of the diagram for mixed states.
  When we say ``a Voronoi diagram with respect to divergence for pure
  states'', it means a diagram obtained by taking a limit of a diagram
  for mixed states.

The following theorem shown in \cite{kato06a} characterize the Voronoi
diagrams that appear on the sphere of one-qubit pure states.
\begin{theorem}
\label{th:coincidence}
For given one-qubit pure states, the following four Voronoi
diagrams are equivalent for pure states:
\begin{enumerate}
\item
the Voronoi diagram with respect to the Fubini-Study distance,
\item
the Voronoi diagram with respect to the Bures distance,
\item
the Voronoi diagram on the sphere with respect to the ordinary
geodetic distance,
\item
the section of the three-dimensional Euclidean Voronoi diagram with the
sphere, and
\item
the Voronoi diagram with respect to the divergences, i.e.~$V_D$ and $V_{D^*}$.
\end{enumerate}
\end{theorem} 
\begin{proof}
 See \cite{kato06a}.
\end{proof}

Note that generally the limit of $V_D$ is meaningless because the
diagram depends on how the sites converges. However, in a one-qubit
system, since there is a special symmetry, we can think of a ``natural''
convergence of the sites. Actually just take the sites on the same
sphere with its center at the origin, and converge the radius of the
sphere to 1. In the theorem above, ``the Voronoi diagram with respect to
$V_D$'' means the limit of the diagram obtained by such a
convergence. Since that definition is only valid for a one-qubit system,
for a higher level, only $V_{D^*}$ is defined.

For mixed states, we found we can say the similar thing. We have
the following theorem:
\begin{theorem}
\label{th:one-qubit-mixed}
 For given one-qubit pure states. the following Voronoi diagrams are
 equivalent for any states (including mixed states)
\begin{itemize}
\item
the Voronoi diagram with respect to the Bures distance,
\item
the Euclidean Voronoi diagram with the sphere, and
\item
the Voronoi diagram with respect to the divergences, i.e.~$V_D$ and $V_{D^*}$.
\end{itemize}
\end{theorem}
\begin{proof}
 See \cite{kato06b}.
\end{proof}

\section{Holevo Capacity for One-qubit Quantum States}
\label{sec:holevo}
Our first motivation to investigate a Voronoi diagram in quantum states
is the numerical calculation of the Holevo capacity for one-qubit
quantum states \cite{oto04}. We explain its method in this section.  In
order to calculate the Holevo capacity, some points are plotted in the
source of channel, and it is assumed that just thinking of the images of
plotted points is enough for approximation. Actually, the Holevo
capacity is reasonably approximated taking the smallest enclosing ball
of the images of the points.  More precisely, the procedure for the
approximation is the following:
\begin{enumerate}
 \item Plot equally distributed points on the Bloch ball which is the
       source of the channel in problem.
 \item Map all the plotted points by the channel.
 \item Compute the smallest enclosing ball of the image with respect to
       the divergence. Its radius is the Holevo capacity.
\end{enumerate}
In this procedure, Step 3 uses a farthest Voronoi diagram. That is the
essential part to make this algorithm effective because Voronoi diagram
is the known fastest tool to seek a center of a smallest enclosing ball
of points.

However, when you think about the effectiveness of this algorithm, there might
arise a question about its reasonableness. Since the Euclidean distance and the
divergence are completely different, Euclideanly uniform points are
not necessarily uniform with respect to the divergence. We gave partial
answer to that problem by Theorem \ref{th:coincidence}. At least, on the
surface of the Bloch ball, the coincidence of Voronoi diagrams implies
that the uniformness of points with respect to Euclidean distance is
equivalent to the uniformness with respect to the divergence.

\section{Voronoi Diagrams for Three or Higher Level Quantum States}
\label{sec:higher-level}

In \cite{kato06a}, the authors showed that the coincidence of the
divergence-Voronoi diagram and the Euclidean Voronoi diagram which
happens in one-qubit case never occurs in a higher level case. In this
section, we explain the outline of the proof described in
\cite{kato06a}.  To show that fact, it is enough to look at some section
of the diagrams with some (general dimensional) plane. If the diagrams
do not coincide in the section, you can say they are different.

Suppose that $d\geq 3$ and that the space of general quantum states is
expressed as (\ref{multi-level-rho}), and let us think the
section of it with a $(d+1)$-plane:
\begin{equation} \label{rep-section}
 \xi_{d+2} = \xi_{d+3} = \cdots = \xi_{d^2-1}.
\end{equation}
Then the section is expressed as:
\begin{equation}
 \label{section-rho}
\rho=
\begin{pmatrix}
  \frac{\xi_1+1}{d} & \frac{\xi_{d}-i\xi_{d+1}}{2}& & & \smash{\lower1.0ex\hbox{\bg 0}}\\
 \frac{\xi_{d}+i\xi_{d+1}}{2} & \frac{\xi_2+1}{d}& & & \\
 & & \ddots & & \\
 & & & \frac{\xi_{d-1}+1}{d}&\\
 \smash{\hbox{\bg 0}} & & & &
 \frac{-\sum_{i=1}^{d-1}\xi_i+1}{d}
\end{pmatrix}
.
\end{equation}
Under that condition, we obtain the the expression of the boundary as follows:
\begin{equation}
 \label{boundary-divergence}
(\eta_d-\tilde{\eta}_d)\xi_d + (\eta_{d+1} - \tilde{\eta}_{d+1})\xi_{d+1}
+\frac{4(\eta_1-\tilde{\eta}_1)\left(\xi_1-\frac{d-2}{2}\right)}{d^2}
=0.
\end{equation}
The detailed process to obtain this equation is described in \cite{kato06a}.

Moreover, (\ref{boundary-divergence}) tells us a
geometric interpretation of this boundary. We obtain the following
theorem:
\begin{theorem}
 On the ellipsoid of the pure states which appears in the section with
 the $(d+1)$-plane defined above, if transfered by a linear transform which
 maps the ellipsoid to a sphere, the Voronoi diagram with respect to the
 divergence coincides with the one with respect to the geodesic
 distance.
\end{theorem}
\begin{proof}
See \cite{kato06a}
\end{proof}

Now we work out the Voronoi diagram with respect to Euclidean
distance. Under the assumption above, the Euclidean distance is
expressed as
\begin{align}
 d(\sigma,\rho)
 &=(\eta_1-\xi_1)^2+\!(\eta_2-\xi_2)^2+
 (\eta_d-\xi_d)^2 + (\eta_{d+1}-\xi_{d+1})^2\nonumber\\
 &=2(\eta_1-\xi_1)^2+
 (\eta_d-\xi_d)^2 + (\eta_{d+1}-\xi_{d+1})^2,
\end{align}
and we get the equation for boundary as
\begin{multline}
  \label{boundary-euclidean}
 d(\sigma,\rho)-d(\tilde{\sigma},\rho)
 =-4(\eta_1-\tilde{\eta}_1)\xi_1
 -2(\eta_d-\tilde{\eta}_d)\xi_d
 -2(\eta_{d+1}-\tilde{\eta}_{d+1})\xi_{d+1}
 +2(\eta_1^2-\tilde{\eta}_1^2)\\
 +(\eta_d^2-\tilde{\eta}_d^2)
 +(\eta_{d+1}^2-\tilde{\eta}_{d+1}^2)=0.
\end{multline}
By comparing the coefficients of $\xi_1$, $\xi_d$, and $\xi_{d+1}$,
we can tell that the boundaries expressed by
(\ref{boundary-divergence}) and (\ref{boundary-euclidean}) are
different.

\section{Bures distance and Fubini-Study Distance}
\label{sec:higher-level-bures}

Although the diagrams with respect to the divergence and the Euclidean
distance are different as shown in the previous section, for the divergence, the Bures-distance and the Fubini-Study
distance, the coincidence of diagrams which holds for one-qubit states
also holds for a higher level. It is stated as follows:
\begin{theorem}
 In a general level quantum system, for pure states,
 the following diagrams are equivalent:
\begin{itemize}
 \item diagram with respect to the divergence
\item diagram with respect to Fubini-Study distance
\item diagram with respect to Bures distance
\end{itemize}
\end{theorem}

The equivalence between the Fubini-Study diagram and the Bures diagram
is obvious because
\begin{equation}
 \label{equivalence_of_equations}
  d_\mathrm{B}(\rho,\sigma)\leq d_\mathrm{B}(\rho,\tilde{\sigma})
\ \Longleftrightarrow\  \Tr\rho\sigma\geq \Tr\rho\tilde{\sigma}
\ \Longleftrightarrow\  
d_\mathrm{FS}(\rho,\sigma)\leq d_\mathrm{FS}(\rho,\tilde{\sigma})
\end{equation}
Hence the only thing to show is the coincidence between the diagram by
Bures distance and the diagram by divergence. The rest of the proof is described in
\cite{kato06b}.

\section{Concluding Remarks}
\label{sec:conclusion} We showed that in the one-qubit system, the Voronoi
diagrams with respect to some distances are the same. Among them,
coincidence of the divergence-Voronoi diagram and the Bures-Voronoi
diagram for pure states especially plays an important role in a numerical
calculation of a capacity of a quantum communication channel.

Our next target is a capacity evaluation of a higher level quantum
communication channel. However, as we showed in this paper, the
theoretical support for the one-qubit numerical capacity estimation ---
the coincidence of the divergence-Voronoi diagram and Bures-Voronoi
diagram --- does not hold in a higher level. On the other hand, we
showed that the divergence-, Bures-, and Fubini-Study-Voronoi diagram
are all the same even for a higher level.

The facts we showed have a negative impact for a higher-level numerical
capacity estimation. The naive extension of the method for one-qubit
quantum states is found to be not efficient for a higher
level. Nevertheless, our geometrical analysis for quantum states has
contributed to a further interpretation of the space of quantum
states. Our next work is a numerical capacity estimation especially for
three-level system, and we believe the analysis in this paper will be
helpful for that objective.

\bibliography{kkato2006}
\bibliographystyle{splncs}

\end{document}